# Strain Influence on Optical Absorption
# of Giant Semiconductor Colloidal Quantum Dots

*Tudor E. Pahomi and Tiberius O. Cheche*[*]

Faculty of Physics, University of Bucharest, RO-077125, Măgurele - Bucharest, Romania, EU

**Abstract.** The lattice mismatch strain field of core/multishell structures with spherical symmetry is modeled by a linear continuum elasticity approach. The effect of the strain on the energy structure and linear optical absorption in large core/shell/shell spherical semiconductor quantum dots is analyzed. Localization of the photoexcited carriers induced by coating is found to play an important role in explaining the optical stability of large CdSe/CdS/ZnS and ZnTe/ZnSe/ZnS quantum dots.

## 1.    Introduction

As 'The Next Big Thing' in photovoltaics [1], the colloidal multishell semiconductor quantum dots (QDs) have led to the development of high-efficiency solar cells. To overcome the crystal irregularities induced by the lattice mismatch in the synthesis of these colloidal nanocrystals, the use of a strain-adapting intermediate shell in core/shell (CS) QDs has been proposed. Thus, 'giant' core/shell/shell (CSS) QDs of 18–19 monolayers shell thickness of are synthesized [2,3]. There are several theoretical studies of multi-component nanocrystals, in which the role of the strain is considered by first-principle calculations, by using, for example, the density-functional tight-binding method [4] or local density approximation [5] or density-functional theory [6]. Unfortunately, limitations of these *ab-initio* calculations (e.g., bandgap underestimation) make difficult comparison of their results with the experiment. More important, the main problem of the first-principle calculations, the computational cost, can make the method inadequate for larger structures, such as the large CSS QDs. On the other hand, the widely used for analyzing the linear elasticity of epitaxial strained heterointerfaces, the valence force field method (see, e.g., Ref. [7]) is dependent on a priori information regarding the interface structure and surface passivation. The continuum elasticity approach in the limits of homogeneous and isotropic materials has been shown to be in good agreement with the valence force field models

---
[*] E-mail: cheche@gate.sinica.edu.tw

Tel.: +40 724 536 908

for semiconductor QDs of spherical shape and cubic symmetry (see, e.g., Ref. [8]). In this context, we propose a continuum elasticity model for the lattice mismatch strain field in such nanocrystals. Keeping justified simplicity, based on our strain field model, we consider a two-band model within the effective mass approximation to theoretically investigate the energy structure and light absorption of a CSS QD with thick shells. In our model, we consider *ideal* multilayer structures. We assume the defects and impurities with low concentration are located at the interfaces, as reported by experiment, (see, e.g., Ref. [9]), and consequently, do not significantly influence the lattice mismatch strain field.

## 2. Theoretical model

### 2.1. *Strain field and the band lineup in the presence of strain*

First, we describe our method for calculus of the lattice mismatch strain field. For spherical core/shell nanocrystals the displacement ($\mathbf{u}$) has radial symmetry, that is the field is *irrotational*, and within the continuum elasticity approach the equilibrium equation is simply: grad div $\mathbf{u} = 0$ [10]. Linear stress ($\sigma_{ij}$)-strain ($\varepsilon_{ij}$) tensor relation is used to obtain the strain field. For a CSS QD with radii $r_1$ (for the core), $r_2$ (for core+middle shell), and $R$ (for the total radius of core+middle shell+outermost shell), we impose the following boundary conditions: (i) continuous stress at the interfaces, (ii) zero pressure outside QD, and (iii) *shrink-fit* induced by the lattices mismatch (which connects the continuum elastic and the discrete crystalline approaches). The corresponding algebraic equations are:

$$\sigma_{rr}^A(r_1) = \sigma_{rr}^B(r_1), \qquad \sigma_{rr}^B(r_2) = \sigma_{rr}^C(r_2), \tag{1a}$$

$$\sigma_{rr}^C(R) = 0, \tag{1b}$$

$$u_r^A(r_1) - u_r^B(r_1) = \varepsilon_1 r_1, \quad u_r^B(r_2) - u_r^C(r_2) = \varepsilon_2 r_2, \tag{1c}$$

where $\varepsilon_1 = (a_B - a_A)/a_A$ and $\varepsilon_2 = (a_C - a_B)/a_B$ are relative lattice mismatches, and $A$, $B$, and $C$ denote the core, middle, and outermost shell, respectively. Detailed expressions of the strain tensor components are presented in Appendix A. When adapting our analytic expressions for two shells to the case of one shell or core embedded in infinite matrix, we recover the results of the previous works, Ref. [11] and Ref. [12], respectively. The method can also be applied to cylindrical multilayer structures. Irrotational displacement field of the form, $\mathbf{u} = (u_r(r), 0, const.)$ can be used to evaluate the strain field in: (i) two-dimensional circular multilayer structure or (ii) three-dimensional multilayer structure by assuming a certain form of the tensor $e_{zz}$ for each component (core, shells) of the structure (see an example in Ref. [11]).

Second, we model the heterostructures band lineup, which is a crucial in obtaining accurate predictions of the energy structure by the effective mass approach. In semiconductor QDs, the lattice-mismatch strain induces deformations that shift both valence band (VB) and conduction band (CB). Thus, the values of the VB and CB extrema at the $\Gamma$ point (we consider direct band semiconductors) are given by the equation [13]:

$$E_{v,c} = E_{v,c}^u + a_{v,c} e_{hyd},\qquad(2)$$

where the unstrained values are related by $E_c^u = E_v^u + E_g$ and $E_g$ is the unstrained bandgap, $a_{v,c}$ is the volume deformation potential (subscript $v$ for VB, $c$ for CB), and $e_{hyd}$ is the hydrostatic strain.

### 2.2. Single particle states

The single particle states are obtained by solving the Schrödinger equation for the envelope wave function, $H\psi(\mathbf{r}) = E\psi(\mathbf{r})$ within the one-band effective Hamiltonian, $H = [2m^*(r)]^{-1} p^2 + V(r)$, where $m^*(r)$ is the photoexcited carrier $r$-dependent effective mass, and $V(r)$ is the step confinement potential generated by the band-offset of the materials in presence of the strain. The solution is a product of radial function and spherical harmonics, $\psi_{nlm}(\mathbf{r}) = R_l(r)Y_l^m(\theta,\varphi)$. The radial solution is a linear combinations of spherical or modified spherical Bessel functions (see Appendix B, Eqs. (B. 1-4)). Imposing the physical conditions of continuity, we obtain the transcendental equation valid for a *general* form of the three-region step confinement potential (see Fig. 1 and details in Appendix B, Table B. 1):

$$\frac{m_C f_1^{B'}(r_2) - m_B F^C f_1^B(r_2)}{m_C f_2^{B'}(r_2) - m_B F^C f_2^B(r_2)} = \frac{m_A f_1^A(r_1) f_1^{B'}(r_1) - m_B f_1^{A'}(r_1) f_1^B(r_1)}{m_A f_1^A(r_1) f_2^{B'}(r_1) - m_B f_1^{A'}(r_1) f_2^B(r_1)},\qquad(3)$$

where $F^C = [f_1^{C'}(r_2) f_2^C(R) - f_1^C(R) f_2^{C'}(r_2)]/[f_1^C(r_2) f_2^C(R) - f_1^C(R) f_2^C(r_2)]$, $f_{1,2}^{A,B,C}$ are Bessel functions given explicitly in Appendix B, Table B.1; the indices $l$ for the order of Bessel functions and star for the effective mass are omitted, and the prime is used to denote the first radial derivative.

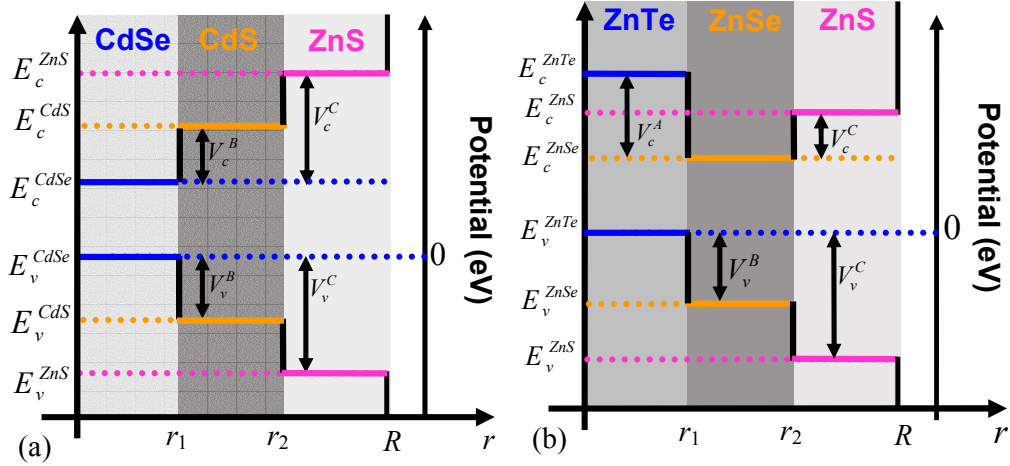

**Fig. 1.** Schematic band lineups in CSS QDs in presence of strain: (a) $V_I(r)$ for CdSe/CdS/ZnS; (b) $V_{II}(r)$ for ZnTe/ZnSe/ZnS.

### 2.3. Excitonic effect

A more accurate description of the optical properties requires estimation of the excitonic effect. This task is more complex as beyond the electron-hole exchange interaction (EHEI) and correlation interactions, the polarization charge induced at the interfaces and the screened dielectric constant should be taken into account. However, the Coulomb electron-hole interaction is usually the leading term of the carrier interaction in excitonic systems. Thus, modeling the excitonic effect by Coulomb electron-hole interaction mediated by a *homogenized* screened dielectric constant is at least satisfactory in estimating the absorption in CSS QDs. We consider a simplified two-band Hamiltonian, by keeping only the kinetic part and Coulomb electron-hole interaction. Neglecting in a first approximation EHEI is adequate. For example, in spherical InAs QD of radius 3nm it is of 2.093meV comparatively to the Coulomb interaction of 60.6meV [14] and of order 0.1 meV in CdSe/CdS QD with thick shell [15]. In our large QDs, the Coulomb interaction is of order of ten meV. We write the exciton state as a configuration interaction expansion, $\left|\Psi\right\rangle = \sum_{i,j=1} C_{ij} c_i^+ h_j^+ \left|0\right\rangle$, with $\left|0\right\rangle$ the ground state (no excited electron or hole particle), and $c_i^+$ ($h_i^+$) creation operator of the electron (hole) state "$i$". With the algebra in the second quantization one obtains the secular equation (Appendix C, Eqs. (C. 10-12)):

$$\sum_{i,j,k,l=1}\left[\left(E_i^e + E_j^h - E\right)\delta_{ik}\delta_{jl} + V_{ijkl}^{eh}\right]C_{kl} = 0,$$ (4)

where $V_{i,j,k,l}^{eh} = -e^2\left(4\pi\varepsilon_0\bar{\varepsilon}_r\right)^{-1}\iint_V d\mathbf{r}_e d\mathbf{r}_h \psi_i^e(\mathbf{r}_e)^* \psi_j^h(\mathbf{r}_h)^* \left|\mathbf{r}_e - \mathbf{r}_h\right|^{-1} \psi_k^h(\mathbf{r}_h) \psi_l^e(\mathbf{r}_e)$ with $\bar{\varepsilon}_r$ the homogenized screened relative dielectric constant.

### 2.4. *Optical absorption*

We find the exciton linear absorption coefficient for a single QD at low temperatures is given by (Appendix C, Eqs. C. (1-9, 14-17)):

$$\alpha_{QD}(\omega) = \frac{\alpha_0}{\omega} \sum_{\tau} \left| \sum_{i,j=1} C_{ij}^{(\tau)} \left\langle \psi_i^e \middle| \psi_j^h \right\rangle \right|^2 \frac{\gamma}{(E_\tau - \hbar\omega)^2 + \gamma^2}, \tag{5}$$

where $\alpha_0$ includes the bulk dependence of the material parameters (see Appendix C, Eq. (C.17)), $E_\tau$ is the energy of the exciton in state $\tau$, and $\gamma$ is the homogeneous electronic broadening. The multishell character is reflected by the excitonic optical matrix element

$$f_{0\tau} = \left| \sum_{i,j=1} C_{ij}^{(\tau)} \left\langle \psi_i^e \middle| \psi_j^h \right\rangle \right|^2 \tag{6}$$

a quantity that can be related to the exciton oscillator strength. As the parameters entering Eq. (5) characterizes the QD, we name it *single* QD absorption coefficient (see Appendix C, Eq. (C.17)). For the absorption in colloidal QD solutions, the measured quantity in experiment, we introduce the absorption coefficient:

$$\alpha_{sol} = -c_{QD}^{1/3} \ln\left[1 - 4R^2 c_{QD}^{2/3} \left(1 - e^{-2R\alpha_{QD}}\right)\right], \tag{7}$$

where $c_{QD}$ is solution concentration. As the parameters entering Eq. (7) characterizes the colloidal QD solutions, we name it *colloidal* absorption coefficient. Derivation is given in Appendix C (Eqs. (B. 18-20)). For $\alpha_{QD}R \ll 1$ (that is QDs with radius smaller than 10nm) and $Rc_{QD}^{1/3} \ll 1$ (that is for dilute solutions)

$$\alpha_{sol} = 8R^3 c_{QD} \alpha_{QD}. \tag{8}$$

Thus, comparing Eq. (5) and Eq. (8), we obtain that $\alpha_{sol}^{(\tau)}$, the colloidal absorption coefficient corresponding to the exciton state $\tau$ at resonance, in limit of dilute solutions, is proportional to $f_{0\tau}$.

## 3. Results and discussion

### 3.1. *Energy structure*

In what follows, we apply the above theory to predict the energy structure and the fundamental excitonic absorption (FEA) of CSS QDs. In principle, the continuum strain approach works for thicker shells, consequently, we model the 'giant' CSS QDs from Refs. [3,16]. We consider spherical type-I CdSe/CdS/ZnS QDs with core radius of 2nm and the middle shell thickness of 11ML (hereafter denoted as $QD^I$) and spherical type-II ZnTe/ZnSe/ZnS QDs with core radius of 2.2nm and the middle shell thickness of 6ML (hereafter denoted as $QD^{II}$). We take in our estimation shell thickness large enough to keep premises of continuum elasticity approach (in the modeling it is at least of 6ML). $QD^I$ is interesting in applications for its optical and chemical stability [2,3], and $QD^{II}$, following the excited charge separation, for its photovoltaic properties [16].

In the first step of application of the theory, we obtain the strain field of the $QD^{I,\,II}$. Calculus shows that the core and middle shell are compressed ($e_{hyd}^{CdSe}$, $e_{hyd}^{ZnTe}$, $e_{hyd}^{CdS}$, $e_{hyd}^{ZnSe} < 0$) and the outermost shell is dilated ($e_{hyd}^{ZnS} > 0$). Then, from Eq. (2), we obtain the band lineups as $V_I(r)$ for $QD^I$ and $V_{II}(r)$ for $QD^{II}$ with the notations from Fig. 1.

In the second step, we characterize the single particle states in a one-band approximation with the potential band offset built with the strain band lineups. In the II-VI semiconductor heterostructures, following the large bandgap, the CB-VB admixture is insignificant. In addition, for such heterostructures, crossing of the heavy and light holes is expected to be obtained beyond the first few excited states [17]. Thus, we first find the form of the spherical Bessel functions $f_{1,2}^{A,B,C}$ for electron and hole confined by the above $V_{I,\,II}$ potentials. The analytical results obtained for the radial functions are presented in Appendix B (Table B.1). Then, we compute the single particle energy of the electron and hole by using in Eq. (3) the appropriate functions $f_{1,2}^{A,B,C}$ for each carrier type. In the approximation of unmixed light and heavy hole states, the bulk heavy-hole $m_{hh}$ and light-hole $m_{lh}$ masses assumed by the parabolic dispersion of the one-band model are [18]: $m^{hh}_{\phantom{hh}}{}^{lh} = m_0 \gamma_1^{-1}[1 \pm (6\gamma_3 + 4\gamma_2)/(5\gamma_1)]^{-1}$, where $\gamma_i$ ($i = 1, 2, 3$) are the Luttinger parameters, and $m_0$ is the electron mass. As the limit of large nanocrystals is envisaged, we consider the bulk values of the material parameters (their values are presented in Appendix B (Table B.2)). For the monolayer thickness, we take the values 0.4nm for CdS [15], 0.33nm for ZnSe [19], and consider 0.33nm for ZnS. In the calculus, we do find the first four hole states are heavy hole states and the fifth is the first light hole energy level, for both $QD^I$ and $QD^{II}$ and the two-band approximation is justified. This characteristic (common to the wide bandgap semiconductor heterostructures) guarantees an accurate prediction of the several first single particle states by a two-band effective mass approach. The results obtained in this applicative step can be summarized are as follows:

(i) Expected *red-shift* with the first shell thickness is found. The energy structure calculation shows the single particle fundamental interband transition (in absence of excitonic effect and for infinite well: $E_{nL}^{\alpha} = \hbar^2 \left(2m_{\alpha}^*\right)^{-1} k_{nL}^2 R^{-2}$ with $k_{n,L}$ the n-th zero of the spherical Bessel function of order $L$, and $\alpha = e, h$ for electron and hole, respectively. For the fundamental state $L=0$, $n=1$.) increases with the coating: from 302nm for CdSe spherical nanocrystal with (core) radius 2nm to 587nm for $QD_0^{I}$ (notation for QD with core radius of 2nm and CdS shell thickness of 11ML); from 320nm for ZnTe spherical nanocrystals with (core) radius 2.2nm to 529nm for $QD_0^{II}$ (notation for QD with core radius of 2.2nm and ZnSe shell thickness of 6ML).

(ii) Expected *slight* variation of the single particle energy with the ZnS shell thickness is obtained. The single particle energies are shown in Fig. 2 for $QD^{I}$ and $QD^{II}$ as function of total radius $R$. We obtain, the single particle fundamental interband transition is blue-shifted when coating, a tendency which asymptotically reduces for both $QD_x^{I}$ (notation for CdSe/11CdS/ZnS with $x$ML ZnS) and $QD_x^{II}$ (notation for ZnTe/6ZnSe/ZnS with $x$ML ZnS). We name the coating with $x = 6\,\text{ML}$ as ZnS *coating* and $x > 6\,\text{ML}$ as ZnS *overcoating*.

(iii) Specific localization of the photoexcited electron and hole is found. Thus, the electron and hole location in the core for $QD^{I}$ and with the hole located in the core and the electron in the shell for $QD^{II}$ is obtained. For a more comprehensive image on the carrier localization, useful in engineering such nanocrystals, in Appendix B (Fig. B.1), we represent the charge density (orbitals) for the two nanocrystals types.

(iv) By varying the shell thickness of both middle and outermost shell, the strain field model predicts that a band lineup that favors the hole escape from the core is not possible for either $QD^{I}$ or $QD^{II}$. Thus, we find that by a mechanism based on the epitaxial strain, in $QD^{II}$ the hole can not be extracted from the core.

It is worth mentioning that in our structures, the interlevel single particle states for energy domain we analyze (see Fig. 2) is larger than 10meV, consequently larger than the estimated EHEI of 0.1meV for large QDs. Thus, the exciton approximation with Coulomb electron-hole interaction as leading term and neglected EHEI is justified. For estimation of FEA, we take the basis set of the configurations corresponding to the first four energy levels obtained from Eq. (3). The configurations have similar structure for type-I and type-II CSS QDs. These configurations are obtained from the $(2L+1)$ degenerated electron and hole states with $(n, L) = (1, 0), (2, 1), (3, 2)$ and the $(4, 3)$ electron states and the $(4, 0)$ hole state. There are 160 configurations obtained by replacing one from the set of (1+3+5+1) VB orbitals with one of the (1+3+5+7) CB orbitals. In the basis we work, the convergence of excitonic ground state ($X_g$) energy is checked.

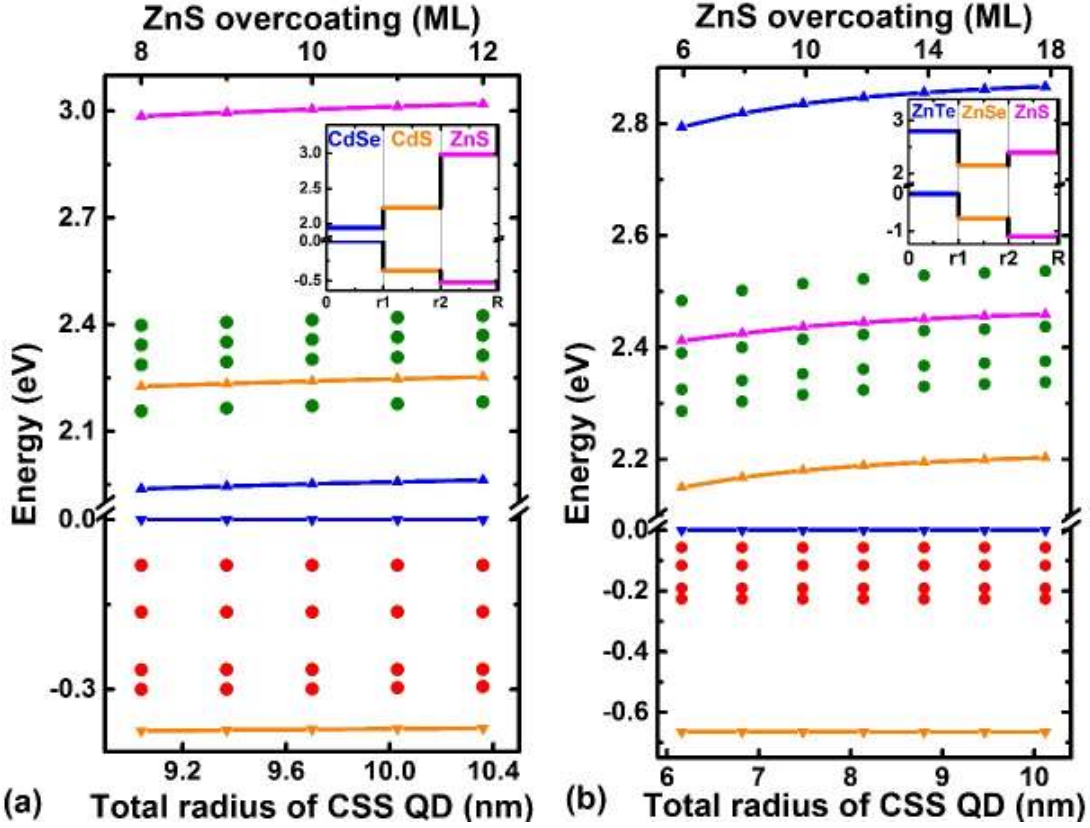

**Fig. 2**. Energy of the first four electron (green color) and hole (red color) single particle states in CSS QDs of total radius $R$: (a) CdSe/11CdS/ZnS QDs; (b) ZnTe/6ZnSe/ZnS QDs. Continuum lines with up (down) triangle symbols show the band lineup in presence of lattice mismatch strain of electron (hole) states for ZnTe-blue color, ZnSe-orange color, ZnS-violet color in figure (a) and for CdSe-blue color, CdS-orange color, ZnS-violet color in figure (b). The insets are for the lineup guidance. Zero reference is ZnTe CB edge, see Fig. 1.

### 3.2. *Excitonic effect and optical absorption*

Next, we consider the excitonic effect, and try to explain the chemical and optical stability reported for CSS QDs. The optical stability manifests by weakly affected absorption and improved fluorescence quantum yield in overcoated or thick shell CS QDs, such as CdSe/CdS [2,3,20], CdZnSe/ZnSe [21], and ZnTe/ZnSe [16]. To evaluate the excitonic effect by our model we need the value of the screened dielectric constant. For estimation, we setup its value from the fit the experimental FEA reports for thick multilayer nanocrystals, namely, Ref. [3] for $QD^I$ and Ref. [16] for $QD^{II}$.

We first consider modeling the optical absorption of $QD^I$. FEA for $QD_0^{Ia}$ (notation for CdSe/CdS QD with core radius of 2nm and CdS shell thickness of 19ML) is about 620nm (Ref. [3]) and it reflects the core absorption. The volume-weighted average of the static dielectric constants of the two components is $\varepsilon_{av}(QD_0^{Ia}) = 8.9$. By our excitonic model for $QD_0^{Ia}$ one obtains the homogenized screened dielectric constant that fits this absorption line is $\overline{\varepsilon}(QD_0^{Ia}) = 7.2$ (smaller in QD than in the corresponding bulk material, in accord with the general agreement; see, e.g. Ref. [22] or Ref. [23]). Next, to analyze the ZnS coating effect in the $QD^I$ absorption, we heuristically estimate the dielectric constant in $QD_x^I$ by $\overline{\varepsilon}(QD_x^I) = \overline{\varepsilon}(QD_0^{Ia})\varepsilon_{av}(QD_x^I)/\varepsilon_{av}(QD_0^{Ia})$ and compute FEA variation with the ZnS shell thickness. Fig. 3a shows that $E_{\text{FEA}}(QD_x^I)$ (FEA of $QD_x^I$) is blue-shifted comparatively to $E_{\text{FEA}}(QD_0^I)$ and asymptotically blue-shifted with the ZnS overcoating. Thus, a relative weak change of $E_{\text{FEA}}(QD_x^I)$ with the ZnS overcoating as reported by experiment [3] is obtained. This behavior is primarily the result of the lattice-mismatch strain and secondly of the excitonic effect. In Fig. 3a, $\alpha_{FEA}(QD_x^I)$ (single QD absorption from Eq. (5) for FEA) obtained within our ideal model is decreasing with either ZnS coating or overcoating. On the other hand, the colloidal absorption coefficient, also shown in Fig. 3a for dilute colloidal QD solutions, is weakly changed by either ZnS coating or overcoating. This is in accord with the reported optical stability of such CSS QD [3]. From Eq. (6) we obtain $f_{0X_g}$ is very slightly varying with either ZnS coating or overcoating.

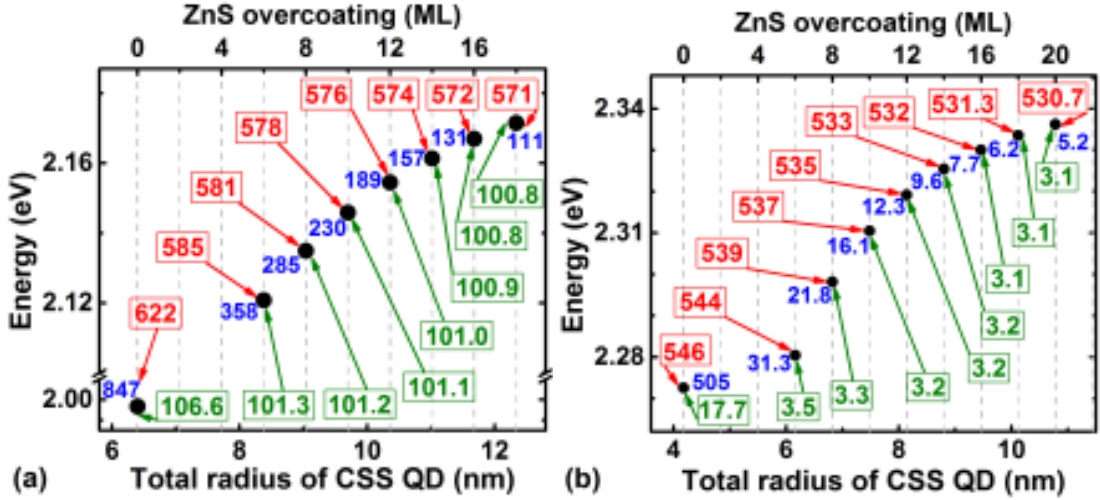

**Fig. 3**. FEA (black circle and red, borderd label for the energy expressed in units of nm), $\alpha_{FEA}$ (blue label, units of $10^3 \, \text{m}^{-1}$), $\alpha_{sol}$ (green, borderd label, units of $\text{m}^{-1}$) for: (a) CdSe/11CdS/xZnS QD; (b) ZnTe/6ZnSe/xZnS QD. $\alpha_{sol}$ is obtained for concentration of $100 \, \mu\text{M}$ for the colloidal QD solution.

The basis set we use also reproduces the absorption line of approximately 500nm associated in Ref. [3] to the CdS bulk bandgap (511nm for $QD_{x=12}^{I}$). As expected, $X_g$ has spherical symmetry, the configuration $c_{100}^+ h_{100}^+ |0\rangle$ corresponding to the *s-s* orbital combination having probability of 0.98.

We continue discussion, and consider modeling the optical absorption of $QD^{II}$. FEA for $QD_0^{IIa}$ (notation for ZnTe/ZnSe QD with core radius of 2.2nm and ZnSe shell thickness of 6ML) is about 570nm [16]. We obtain the homogenized screened dielectric constant that fits this absorption line is $\bar{\varepsilon}(QD_0^{IIa}) = 7$. The volume-weighted average of the dielectric constants of the two components is $\varepsilon_{av}(QD_0^{IIa}) = 8.9$. For $QD_x^{II}$, we consider again $\bar{\varepsilon}(QD_x^{II}) = \bar{\varepsilon}(QD_0^{IIa}) \varepsilon_{av}(QD_x^{II}) / \varepsilon_{av}(QD_0^{IIa})$ and analyze the ZnS shell thickness effect on FEA. Fig. 3b shows $E_{\text{FEA}}(QD_x^{II})$ is blue-shifted comparatively to $E_{\text{FEA}}(QD_0^{II})$, and asymptotically blue-shifted with the ZnS overcoating. Thus, our modeling confirms that, as a result of the lattice-mismatch strain and excitonic effect, as in the $QD_x^{I}$ case, the influence of ZnS overcoating on FEA is weak, as reported by experiment [16]. Differently from the $QD_x^{I}$ case, the single QD absorption coefficient is decreasing by one order of magnitude with ZnS coating and less by overcoating (see Fig. 3b). This is the result of the orbitals overlap decreasing with the ZnS coating or overcoating. We predict $\alpha_{sol}$ of dilute solutions also strongly decreases with the ZnS coating, but as in the type-I case it is almost constant with the ZnS overcoating, see Fig. 3b. This dependence can be related to the protecting effect of the ZnS overcoating reported by experiment [16]. From Eq. (6), we obtain $f_{0X_g}$ decreases about 5 times by ZnS coating and is weakly sensitive with the ZnS overcoating. As expected, similarly to the type-I CSS QD, $X_g$ has spherical symmetry, the configuration $c_{100}^{+}h_{100}^{+}|0\rangle$ corresponding to the *s-s* orbital combination having probability of 0.98, with the photoexcited electron located in the middle ZnSe shell and hole located in the core.

By our model, we can obtain useful information about the photoexcited carriers localization and optical absorption characteristics. The results regarding photoexcited carriers localization are presented in Appendix B (Table B.3). In Fig. 4, we also represented the radial distribution probability density of the photoexcited charges in the excitonic ground state (see details in Appendix B, Eqs. (B. 5-6)). We find the ZnS either coating or overcoating effect on radius expectation value is not significant for $QD^{I}$ (ZnS coating weakly moves the electron to the center of QD). On the other hand, for $QD^{II}$, the hole radius expectation value is practically not affected by the ZnS either coating or overcoating, but the ZnS coating or overcoating has a strong effect on the electron localization, namely, the electron moves to the middle of the ZnSe shell. Thus, according to our model in both $QD^{I}$ and $QD^{II}$ the electron and hole is not confined in the proximity of the interfaces and surface.

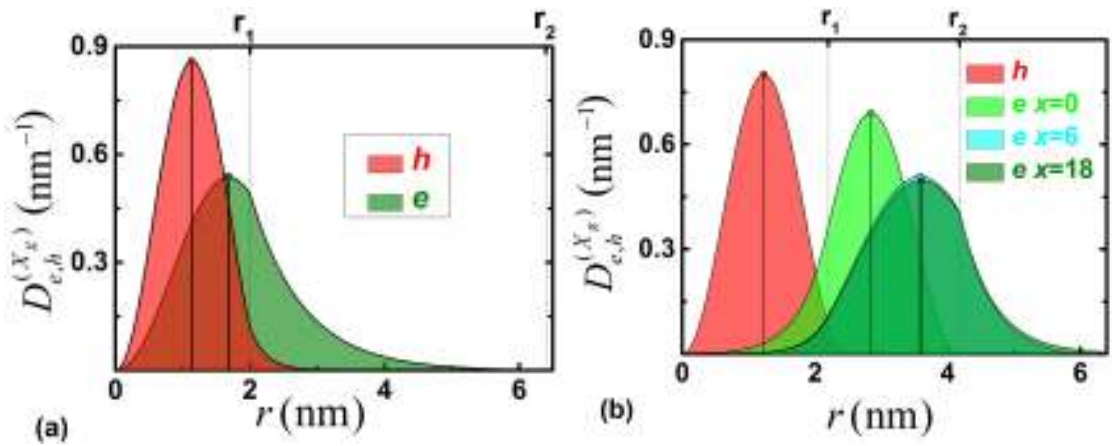

**Fig. 4**. Photoexcited carriers distribution probability density for electron (*e*) and hole (*h*): (a) CdSe/11CdS/*x*ZnS CSS QD; (b) ZnTe/6ZnSe/*x*ZnS CSS QD. Hole is localized in the core while the electron is localized in the core for type-I CSS QD and in the middle shell for type-II CSS QD. Current continuity is apparent in the represented distribution probability.

*3.3. Emission considerations*

Emission modeling is a more complex process. The only emission related quantity found in our model, $f_{0X_g}$ (proportional to the rate of radiative exciton recombination by spontaneous emission, see, e.g., Ref. [24]) is practically not changed by either ZnS coating or overcoating. If this finding could explain the stable emission intensity observed in ultra-thick-shell CdSe/CdS [20], it cannot explain the blinking suppression induced by coating in such nanocrystals. The surface and interface impurities and interface defects and, in addition, very likely as result of the synthesis, the irregular thickness of the shells can not be neglected any longer. They act as trapping states of the photoexcited carriers. Several mechanisms explain blinking in QDs by competition between the intercept of the photoexcited carrier by trapping states and its relaxation to the band-edge [25,26]. Thus, the luminescence is on/off when: (i) by Auger recombination of the photoexcited carriers on the trapping states the QD becomes neutral/charged [27] or (ii) charge fluctuations make the trapping sites inactive/active [20,28]. The on/off dynamics is explained by statistic models, such as the spectral diffusion models [29]. Relevant for the present work is that in the experiment one observes the light intensity fluctuation in CS QDs is reduced with the shell thickness [20]. According to the earlier mentioned localization of the photoexcited carriers in the $QD^I$ with the ZnS coating, our model best fits the model assuming a tunneling barrier between the photoexcited carrier and the trap states, similar to that proposed in Refs. [28] and [30]. QD core coating induces larger photoexcited carrier-trap separation and determines lowering of the trapping probability. In QDs with thicker shell(s), the tunneling is less probable and a continuous luminescence (nonblinking) is expected under continuous photoexcitation of QDs. Thickness irregularity of the shell(s) obtained as the chemical synthesis result can be a physical factor that induces the random ionization and neutralization of the trapping states.

## 4. Conclusions

By this investigation, we developed a continuum elasticity model of the strain field for isotropic, homogeneous and finite size multilayer structures that is applied to spherical large CSS QD nanocrystals. The quantum treatment based on the proposed strain field model can explain the optical stability of the overcoated core-shell QDs. According to our estimations, the measured absorption coefficient in colloidal QD solutions is practically not sensitive with the overcoating for the core-shell QDs analyzed. The most important finding is that related to the photoexcited carriers localization. According to our predictions, in both $QD^{I, II}$ with thick shells we analyzed, the photoexcited carriers are moved away from the surface and interfaces. The hole is strongly confined in the core. The electron is less confined than the hole in the core of the $QD^I$, but it is strictly localized in the middle shell of the $QD^{II}$ we discussed. This shielding of the photoexcited carriers from the surface and interfaces plays an important role in explaining the nonradiative recombination in core-shell(s) QDs. Thus, our model advocates the nonblinking of thick shell QDs is the result of low tunneling rate of the barriers created by the surface or interface located traps, which would lead to lower Auger recombination.

In essence, our continuum model of the strain field for homogenous and isotropic elastic multilayer structures, when implemented in the specific quantum mechanics of the multishell semiconductor nanocrystals, is able to predict the main characteristics of fundamental absorption in the thick shells $QD^{I, II}$ we considered. We believe it is a useful framework, in which improved modeling (multi-band treatment, multi-exciton generation, electron-phonon interaction, statistics blinking and relaxation dynamics consideration) can overcome the present computational limits of the first-principle calculations for more accurate description of the complex optical processes in multilayer QDs.

**Acknowledgements**. Thanks are due to Yia-Chung Chang for useful discussions.

**Appendix A: Strain field**

Displacement field in spherical coordinates is of the form $\mathbf{u} = (u_r(r), 0, 0)$, and consequently it is irrotational. To find the strain tensor from the equilibrium equation $\mathrm{grad\,div}\,\mathbf{u} = 0$ [1] for spherical multilayer structures within the continuum elasticity model, we seek solutions of the form $u_r^X(r) = X_1 r + X_2 / r^2$ with: $X = A$, $X_1 = A_1$, $X_2 = A_2$=0 for core, $X = B$, $X_1 = B_1$, $X_2 = B_2$ for the first shell, $X = C$, $X_1 = C_1$, $X_2 = C_2$ for the second shell, etc. With Eqs. (1) from the main text, we compute the strain tensor components from the above form of $u_r^X(r)$ and then find the stress tensor $\sigma_{ij}$ by applying Hooke's law [1], $\sigma_{ij} = E(1+\nu)^{-1}\left[e_{ij} + \nu(1-2\nu)^{-1} e_{ll}\delta_{ij}\right]$ (where $E$ is the Young modulus and $\nu$ is Poisson ratio). For two shell spherical structures, we obtain the following non-zero components of the strain tensor. For the core:

$$e_{rr}^A = e_{\theta\theta}^A = e_{\varphi\varphi}^A = \frac{\varepsilon_1 S_1^A + \varepsilon_2 S_2^A}{S_1^A + S_3^A}, \tag{A.1}$$

$$e_{hyd}^A = 3\frac{\varepsilon_1 S_1^A + \varepsilon_2 S_2^A}{S_1^A + S_3^A}, \tag{A.2}$$

where $S_1^A = \dfrac{E_B}{E_C}\dfrac{r_1^3/r_2^3 - 1}{R^3/r_2^3 - 1} - \dfrac{2(1-2\nu_B) + (1+\nu_B)r_1^3/r_2^3}{2(1-2\nu_C) + (1+\nu_C)R^3/r_2^3}$, $\quad S_2^A = -\dfrac{3(1-\nu_B)}{2(1-2\nu_C) + (1+\nu_C)R^3/r_2^3}$, and

$$S_3^A = \frac{E_A}{1-2\nu_A}\left(\frac{1}{E_B}\frac{(1-2\nu_B)(1+\nu_B)\left(r_1^3/r_2^3 - 1\right)}{2(1-2\nu_C) + (1+\nu_C)R^3/r_2^3} - \frac{1}{E_C}\frac{(1+\nu_B)/2 + (1-2\nu_B)r_1^3/r_2^3}{R^3/r_2^3 - 1}\right).$$

For the middle shell:

$$e_{rr}^B(r) = \frac{\varepsilon_1 S_{1*}^B r_1^3/r_2^3 + \varepsilon_2 S_{2*}^B}{S_3^B} + 2\frac{r_1^3}{r^3}\frac{\varepsilon_1(1+S_1^B) + \varepsilon_2(1+S_2^B)}{S_3^B}, \tag{A.3}$$

$$e_{\theta\theta}^B(r) = e_{\varphi\varphi}^B(r) = \frac{\varepsilon_1 S_{1*}^B r_1^3/r_2^3 + \varepsilon_2 S_{2*}^B}{S_3^B} - \frac{r_1^3}{r^3}\frac{\varepsilon_1(1+S_1^B) + \varepsilon_2(1+S_2^B)}{S_3^B}, \tag{A.4}$$

$$e_{hyd}^B = 3\frac{\varepsilon_1 S_{1*}^B r_1^3/r_2^3 + \varepsilon_2 S_{2*}^B}{S_3^B}, \tag{A.5}$$

where $S_1^B = \dfrac{E_B}{E_C}\dfrac{1-2\nu_C + \frac{1}{2}(1+\nu_C)R^3/r_2^3}{(1-2\nu_B)\left(R^3/r_2^3 - 1\right)}$, $S_2^B = -\dfrac{E_B}{E_A}\dfrac{1-2\nu_A}{1-2\nu_B}$, $S_{1*,2*}^B = 1 - 2\dfrac{1-2\nu_B}{1+\nu_B}S_{1,2}^B$, and

$$S_3^B = 1 - \frac{r_1^3}{r_2^3} + S_1^B\left(1 + 2\frac{1-2\nu_B}{1+\nu_B}\frac{r_1^3}{r_2^3}\right) - S_2^B\left[\frac{r_1^3}{r_2^3} + 2\frac{1-2\nu_B}{1+\nu_B}\left(1 - S_1^B\left(\frac{r_1^3}{r_2^3} - 1\right)\right)\right].$$

For the outermost shell:

$$e_{rr}^C(r) = -\frac{\varepsilon_1 S_1^C + \varepsilon_2 S_2^C}{S_4^C}\left(1 - \frac{1+\nu_C}{1-2\nu_C}\frac{R^3}{r^3}\right), \tag{A.6}$$

$$e_{\theta\theta}^C(r) = e_{\varphi\varphi}^C(r) = -\frac{\varepsilon_1 S_1^C + \varepsilon_2 S_2^C}{S_4^C}\left(1 + \frac{1}{2}\frac{1+\nu_C}{1-2\nu_C}\frac{R^3}{r^3}\right), \tag{A.7}$$

$$e_{hyd}^C = -3\frac{\varepsilon_1 S_1^C + \varepsilon_2 S_2^C}{S_4^C}, \tag{A.8}$$

where $\qquad S_1^C = 3\frac{r_1^3}{r_2^3}\frac{1-\nu_B}{1+\nu_B+2(1-2\nu_B)\,r_1^3/r_2^3}, \qquad S_2^C = 1 - 2\frac{E_B}{E_A}\frac{(1-2\nu_A)(r_1^3/r_2^3-1)}{1+\nu_B+2(1-2\nu_B)\,r_1^3/r_2^3},$

$S_3^C = \frac{1}{2}\frac{1+\nu_C}{1-2\nu_C}\frac{R^3}{r_2^3}+1, \;\; S_4^C = S_3^C + \frac{1-S_2^C}{2}\left[\frac{E_C}{E_B}\frac{R^3/r_2^3-1}{1-2\nu_C}\left(\frac{(1+\nu_B)r_1^3/r_2^3+2(1-2\nu_B)}{r_1^3/r_2^3-1}-\frac{E_A}{E_B}\frac{(1+\nu_B)(1-2\nu_B)}{1-2\nu_A}\right)-2S_3^C\right].$

If the elastic constants are the same, $\nu_A = \nu_B = \nu_C = \nu$, $E_A = E_B = E_C$, Eqs. (A.1-8) become:

$$e_{rr}^A = e_{\theta\theta}^A = e_{\varphi\varphi}^A = \frac{2}{3}\frac{1-2\nu}{1-\nu}\left[\varepsilon_1\left(1-r_1^3/R^3\right)+\varepsilon_2\left(1-r_2^3/R^3\right)\right], \tag{A.9}$$

$$e_{hyd}^A = 2\frac{1-2\nu}{1-\nu}\left[\varepsilon_1\left(1-r_1^3/R^3\right)+\varepsilon_2\left(1-r_2^3/R^3\right)\right], \tag{A.10}$$

$$e_{rr}^B(r) = \frac{2}{3}\frac{1-2\nu}{1-\nu}\left(\varepsilon_1\frac{r_1^3}{R^3}\left(\frac{R^3}{r^3}\frac{1+\nu}{1-2\nu}-1\right)+\varepsilon_2\left(1-\frac{r_2^3}{R^3}\right)\right), \tag{A.11}$$

$$e_{\theta\theta}^B(r) = e_{\varphi\varphi}^B(r) = \frac{2}{3}\frac{1-2\nu}{1-\nu}\left(-\varepsilon_1\frac{r_1^3}{R^3}\left(\frac{1}{2}\frac{R^3}{r^3}\frac{1+\nu}{1-2\nu}+1\right)+\varepsilon_2\left(1-\frac{r_2^3}{R^3}\right)\right), \tag{A.12}$$

$$e_{hyd}^B = 2\frac{1-2\nu}{1-\nu}\left(-\varepsilon_1\frac{r_1^3}{R^3}+\varepsilon_2\left(1-\frac{r_2^3}{R^3}\right)\right), \tag{A.13}$$

$$e_{rr}^C(r) = \frac{2}{3}\frac{1-2\nu}{1-\nu}\left(\frac{R^3}{r^3}\frac{1+\nu}{1-2\nu}-1\right)\left(\varepsilon_1\frac{r_1^3}{R^3}+\varepsilon_2\frac{r_2^3}{R^3}\right), \tag{A.14}$$

$$e_{\theta\theta}^C(r) = e_{\varphi\varphi}^C(r) = -\frac{2}{3}\frac{1-2\nu}{1-\nu}\left(\frac{1}{2}\frac{R^3}{r^3}\frac{1+\nu}{1-2\nu}+1\right)\left(\varepsilon_1\frac{r_1^3}{R^3}+\varepsilon_2\frac{r_2^3}{R^3}\right), \tag{A.14}$$

$$e_{hyd}^C = -2\frac{1-2\nu}{1-\nu}\left(\varepsilon_1\frac{r_1^3}{R^3}+\varepsilon_2\frac{r_2^3}{R^3}\right). \tag{A.15}$$

When our expressions for two shells are adapted to the case of one shell or core embedded in infinite matrix, we recover the results of previous works, Ref. [2] and Ref. [3], respectively. If the core (or CS QD) is embedded in infinite matrix we get that the matrix is not strained. Thus, from Eq. (A.13), if $r_2 = R \rightarrow \infty$ one obtains $e_{hyd}^B \rightarrow 0$ or from Eq (A.15) if $R \rightarrow \infty$ one obtains $e_{hyd}^C \rightarrow 0$.

For the material parameters from Table B.2, the relative lattice mismatches are large as follows:

$\varepsilon_{1I} = (a_{CdS} - a_{CdSe})/a_{CdSe} = -0.038$, $\qquad\qquad$ $\varepsilon_{2I} = (a_{ZnS} - a_{ZnSe})/a_{ZnSe} = -0.072$,

$\varepsilon_{1II} = (a_{ZnSe} - a_{ZnTe})/a_{ZnTe} = -0.071$, $\varepsilon_{2II} = (a_{ZnS} - a_{ZnSe})/a_{ZnSe} = -0.044$.

## Appendix B: One-band Schrödinger equation

Schrödinger equation for the envelope wave function within the one-band effective Hamiltonian is

$$\left[\frac{p^2}{2m(r)} + V(r)\right]\psi(\mathbf{r}) = E\,\psi(\mathbf{r}), \tag{B.1}$$

where $m(r)$ is the carrier $r$-dependent effective mass. The solution separates, $\psi_{nlm}(\mathbf{r}) = R_l(r)Y_l^m(\theta,\varphi)$ and the radial one-particle Schrödinger equation reads:

$$\frac{d}{dr}\left(r^2\frac{dR_l}{dr}\right) - l(l+1)R_l + \frac{2m(r)}{\hbar^2}\left[E - V(r)\right]r^2 R_l = 0. \tag{B.2}$$

To solve Eq. (B.2), we introduce the notations, $\rho = k_i \cdot r$ and $R_l(r) = v_l(\rho)$, where $k_i^2 = 2m_i|E - V_i|/\hbar^2$ with index $i$ standing for the materials $A$ (core), $B$ (first shell), $C$ (second shell), in the separate cases of electron and hole. Handling the substitution, Eq. (B.2) reduces to the spherical Bessel differential equation:

$$\rho^2\frac{d^2 v_l(\rho)}{d\rho^2} + 2\rho\frac{dv_l(\rho)}{d\rho} \pm \left[\rho^2 \mp l(l+1)\right]v_l(\rho) = 0. \tag{B.3}$$

The upper sign corresponds to $E > V_i$ and the lower to $E < V_i$, and the solutions of Eq. (B.3) are spherical Bessel functions $j_l(\rho), y_l(\rho)$ or modified spherical Bessel functions $i_l(\rho), k_l(\rho)$, respectively. The radial wave function for the three regions corresponding to core-$A$ $(0 \leq r < r_1)$, middle shell-$B$ $(r_1 \leq r < r_2)$, outermost shell-$C$ $(r_2 \leq r < R)$, are linear combinations of these functions as follows:

$$\begin{cases} R_l^A(r) = A_1^l\,f_1^A(r),\ 0 \leq r < r_1 \\ R_l^B(r) = B_1^l\,f_1^B(r) + B_2^l\,f_2^B(r),\ r_1 \leq r < r_2 \\ R_l^C(r) = C_1^l\,f_1^C(r) + C_2^l\,f_2^C(r),\ r_2 \leq r < R \end{cases} \tag{B.4}$$

with $A_1^l, B_{1,2}^l, C_{1,2}^l$ constants and $f_{1,2}^\eta$ (the index $l$ is omitted here) are spherical Bessel functions with the argument dependent of $k_\eta = \hbar^{-1}\sqrt{2m_\eta|E - V_\eta|}$ (the superscript $\eta$ stands for $A$, $B$, $C$, and the indexing quantum number $l$ is omitted for the $f$ functions). They are given explicitly in Table B.1, for both electrons and holes for the two types of core/shell/shell (CSS) of quantum dots (QDs) we considered. Imposing the physical conditions of continuity, $R_l^A(r_1) = R_l^B(r_1)$, $R_l^B(r_2) = R_l^C(r_2)$, $R_l^c(R) = 0$, and conservation of the probability current, $m_A^{-1}(dR_l^A/dr)_{r \to r_1} = m_B^{-1}(dR_l^B/dr)_{r \to r_1}$, $m_B^{-1}(dR_l^B/dr)_{r \to r_2} = m_C^{-1}(dR_l^B/dr)_{r \to r_2}$, we obtain the transcendental equation valid for a *general* form of the three-region step confinement potential (see main text, Eq. (3)). From condition of normalization $\int_0^R R_l^2(r) r^2 \, dr = 1$, we find numerically $A_1^l, B_{1,2}^l, C_{1,2}^l$ and consequently the explicit analytical expressions for the normalized eigenfunctions.

**Table B.1** The explicit spherical Bessel functions $f_{1,2}^{A,B,C}$ (the index $l$ omitted in notation) for electron and hole in each region according to Eq. (3) from the main text for the two types of CSS QDs considered, CdSe/CdS/ZnS and ZnTe/ZnSe/ZnS.

| | $A(0 \leq r < r_1)$ | $B(r_1 \leq r < r_2)$ | $C(r_2 \leq r < R)$ |
|---|---|---|---|
| CdSe/CdS/ZnS <br><br> Electron or hole | $f_1^A = j_l(k_A r)$ | $f_1^B = \begin{cases} j_l(k_B r), E > V_{c,v}^B \\ i_l(k_B r), 0 < E < V_{c,v}^B \end{cases}$ <br><br> $f_2^B = \begin{cases} y_l(k_B r), E > V_{c,v}^B \\ k_l(k_B r), 0 < E < V_{c,v}^B \end{cases}$ | $f_1^C = \begin{cases} j_l(k_C r), E > V_{c,v}^C \\ i_l(k_C r), 0 < E < V_{c,v}^C \end{cases}$ <br><br> $f_2^C = \begin{cases} y_l(k_C r), E > V_{c,v}^C \\ k_l(k_C r), 0 < E < V_{c,v}^C \end{cases}$ |
| ZnTe/ZnSe/ZnS <br><br> Electron | $f_1^A = \begin{cases} j_l(k_A r), E > V_c^A \\ i_l(k_A r), 0 < E < V_c^A \end{cases}$ | $f_1^B = j_l(k_B r)$ <br><br> $f_2^B = y_l(k_B r)$ | $f_1^C = \begin{cases} j_l(k_C r), E > V_c^C \\ i_l(k_C r), 0 < E < V_c^C \end{cases}$ <br><br> $f_2^C = \begin{cases} y_l(k_C r), E > V_c^C \\ k_l(k_C r), 0 < E < V_c^C \end{cases}$ |
| Hole | $f_1^A = j_l(k_A r)$ | $f_1^B = \begin{cases} j_l(k_B r), E > V_v^B \\ i_l(k_B r), 0 < E < V_v^B \end{cases}$ <br><br> $f_2^B = \begin{cases} y_l(k_B r), E > V_v^B \\ k_l(k_B r), 0 < E < V_v^B \end{cases}$ | $f_1^C = \begin{cases} j_l(k_C r), E > V_v^C \\ i_l(k_C r), 0 < E < V_v^C \end{cases}$ <br><br> $f_2^C = \begin{cases} y_l(k_C r), E > V_v^C \\ k_l(k_C r), 0 < E < V_v^C \end{cases}$ |

In Fig. B.1, we represent the charge density (orbitals) for the two types of nanocrystal. We can see in ZnTe/6ZnSe/ZnS the carrier separation is enhanced. The shape of the orbitals is drawn according to the *z* chosen quantization axis.

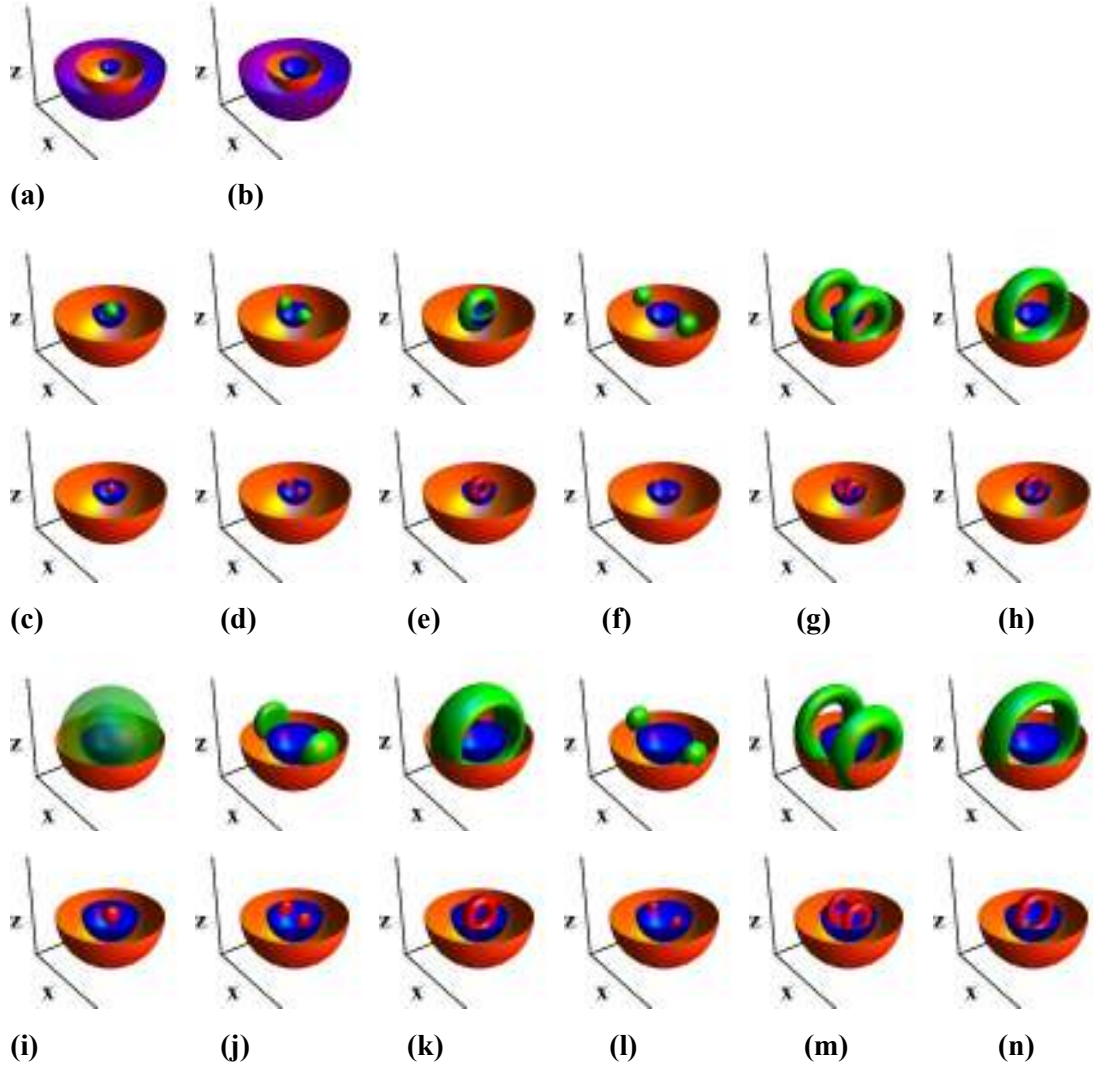

**(a)**   **(b)**

**(c)**   **(d)**   **(e)**   **(f)**   **(g)**   **(h)**

**(i)**   **(j)**   **(k)**   **(l)**   **(m)**   **(n)**

**Fig. B.1.** Electron (green color) and hole (red color) probability density $0.75 \times \left| \psi_{nlm}(\mathbf{r}) \right|^2$ of single particle states for type-I CdSe/11CdS/12ZnS (c-h) and type-II ZnTe/6ZnSe/12ZnS (i-n). (a) and (b) show the nanocrystals without orbitals for the type-I and type-II CSSQDs, respectively. The figures are denoted according to the following quantum numbers: (c, i) $n$=1, $l$=0, $m$=0 (d, j) $n$=2, $l$=1, $m$=0 (e, k) $n$=2, $l$=1, $m$=±1 (f, l) $n$=3, $l$=2, $m$=0 (g, m) $n$=3, $l$=2, $m$=±1 (h, n) $n$=3, $l$=2, $m$=±2. The outermost shell is not represented the figures (c-n) for a better view of the orbitals.

**Table B.2.** Material parameters used in the work.

| | ZnTe | ZnSe | CdSe | CdS | ZnS |
|---|---|---|---|---|---|
| $a$ (Å) | 6.08[a] | 5.65[a] | 6.05[b] | 5.82[b] | 5.40[c] |
| $E$ ($10^{10}\,\mathrm{Nm^{-2}}$) | 4.17[a] | 4.51[a] | 2.87[d] | 3.26[d] | 5.55[c] |
| $\nu$ | 0.363[a] | 0.376[a] | 0.408[d] | 0.410[d] | 0.384[c] |
| $E_{gap}$ (eV) | 2.25[e] | 2.69[e] | 1.74[e] | 2.49[e] | 3.61[e] |
| $E_v$ (eV) | -5.34[e] | -6.07[e] | -6.00[e] | -6.42[e] | -6.6[e] |
| $a_v$ (eV) | 0.79[f] | 1.65[f] | 0.9[b] | 0.4[b] | 2.31[f] |
| $a_c$ (eV) | -5.83[f] | -4.17[f] | -2.00[b] | -2.54[b] | -4.09[f] |
| $\gamma_1$ | 3.74[g] | 3.77[g] | 3.33[h] | 4.11[d] | 2.54[g] |
| $\gamma_2$ | 1.07[g] | 1.24[g] | 1.11[h] | 0.77[d] | 0.75[g] |
| $\gamma_3$ | 1.64[g] | 1.67[g] | 1.11[h] | 1.53[d] | 1.09[g] |
| $m^{lh\ i}$ | 0.152 | 0.148 | 0.18 | 0.15 | 0.225 |
| $m^{hh\ i}$ | 1.092 | 1.292 | 0.90 | 0.60 | 1.582 |
| $m^{ei}$ | 0.20[j] | 0.21[j] | 0.15[k] | 0.22[k] | 0.34[n] |
| $\varepsilon^p$ | 7.4 | 9.1 | 10 | 8.9 | 9 |

[a]Ref. [4]; [b]Ref. [5]; [c]Ref. [6]; [d]Ref. [7]; [e]Ref. [8]; [f]Ref. [9]; [g]Ref. [10]; [h]Ref. [11]; [i]calculated with $m^{lh}_{hh} = m_0\gamma_1^{-1}\left[1 \pm \left(6\gamma_3 + 4\gamma_2\right)/\left(5\gamma_1\right)\right]^{-1}$ from Ref. [12]; [j]Ref. [13]; [k]Ref. [14]; [n]Ref. [15]; [p]Ref. [16].

Regarding the radial distribution probability density of the photoexcited charges, we have $D_\alpha^{(X_g)}(r) = \sum_{i,j=1}\left|C_{ij}^{(X_g)}\right|^2 r^2 \int \left|\psi_i^\alpha(\mathbf{r})\right|^2 d\Omega$ (integration is over the solid angle, and $\alpha = e,\ h$). One observes that:

$$\int_0^R D_\alpha^{(X_g)}(r)\,dr = 1, \tag{B.5}$$

as $\sum_{i,j=1}\left|C_{ij}^{(X_g)}\right|^2 = 1$ from the orthonormalization; $\alpha = e,\ h$, and $C_{ij}^{(X_g)}$ are expansion coefficients of the exciton ground state, $X_g$. The radius expectation value of the photoexcited electron and hole is obtained with

$$\overline{r_\alpha^{(X_g)}} = \sum_{i,j=1}\left|C_{ij}^{(X_g)}\right|^2 \left\langle \psi_i^\alpha \left|r\right| \psi_i^\alpha \right\rangle, \tag{B.6}$$

where $\psi_i^\alpha$ are the envelope wave functions, and $\alpha = e,\ h$.

**Table B.3** Radius expectation value of the electron, $\overline{r_e^{(X_g)}}$, and hole, $\overline{r_h^{(X_g)}}$, for type-I CdSe/CdS/ZnS QD and type-II ZnTe/ZnSe/ZnS QD in the ground excitonic state, expressed in nm.

| $x$ (ML) | | 0 | 6 | 12 | 18 |
|---|---|---|---|---|---|
| Type-I | $\overline{r_h^{(X_g)}}$ | 1.146 | 1.152 | 1.155 | 1.156 |
| | $\overline{r_e^{(X_g)}}$ | 1.899 | 1.890 | 1.878 | 1.873 |
| Type-II | $\overline{r_h^{(X_g)}}$ | 1.202 | 1.214 | 1.214 | 1.215 |
| | $\overline{r_e^{(X_g)}}$ | 2.700 | 3.463 | 3.522 | 3.527 |

**Appendix C: Linear absorption coefficient for single quantum dot**

In the following derivation, we approximate the homogenous medium as formed by QDs in contact and assume each QD is an absorber of volume $V_{QD}$. At the low power densities of the field the assumption of a linear relation between the polarization and the electric field is a good approximation for the description of excitonic optical absorption. It can be obtained using either first order complex susceptibility or Fermi's golden rule, by treating the QD-field interaction as a perturbation. Next, we consider Fermi's golden rule treatment for the excitonic absorption. In macroscopic nonconducting media the gradient of the energy density of monochromatic electromagnetic wave $\mathbf{E} = \mathbf{E}_0 \cos(\omega t - \mathbf{k} \cdot \mathbf{r})$, propagating in $z$ direction is (see. e.g., Ref. [17]):

$$\frac{dw}{dz} = \frac{dw}{dt}\frac{n}{c}. \tag{C.1}$$

The loss of energy per time unit can be written by considering the loss of energy in a *single* QD as

$$\frac{dw}{dt} = -\frac{\hbar\omega}{V_{QD}}R, \tag{C.2}$$

where $R$ is the rate of change of the number of photons. From Eqs. (C.1, 2), one obtains

$$\frac{dw}{dz} = -\hbar\omega R\frac{n}{V_{QD}c}. \tag{C.3}$$

The energy density is $w = \varepsilon_r\varepsilon_0 E_0^2/2$ (Eq. (7.14) in Ref. [18]), then the intensity of the electromagnetic wave is (Eq. (7.13) in Ref. [18])

$$I = \sqrt{\frac{\varepsilon_r\varepsilon_0}{\mu_r\mu_0}}\frac{E_0^2}{2} \approx c\varepsilon_0\varepsilon_r\frac{E_0^2}{2n} = \frac{wc}{n}, \tag{C.4}$$

where we considered $n = \sqrt{\mu_r\varepsilon_r} \cong \sqrt{\varepsilon_r}$ (Eq. (7.5) in Ref. [18]). From Eq. (C.3) and the derivative of Eq. (C.4) one obtains

$$\frac{dI}{dz} = \frac{c}{n}\frac{dw}{dz} = -\frac{\hbar\omega R}{V_{QD}}.$$ (C.5)

With the Beer-Lambert law, $dI/dz = -\alpha_{QD}I$, from Eq. (C.4) and Eq. (C.5), we have for the *single* QD absorption coefficient

$$\alpha_{QD} = \frac{2\hbar\omega}{cn\varepsilon_0 E_0^2 V_{QD}} R.$$ (C.6)

Expression for $R$ can be obtained following the standard textbook derivation of the Fermi golden rule. Next, for the safety of correct factors of the Dirac delta function in the expression of single QD absorption coefficient, we point out the steps of this derivation. Thus, the semi-classical QD-field interaction is written as

$$H_{XF}(t) = \frac{e}{m_0}\mathbf{A}(t)\cdot\mathbf{P} = -\frac{i\,e\,E_0}{2\,m_0\,\omega}\left(e^{-i\omega t} - e^{i\omega t}\right)\mathbf{e}\cdot\mathbf{P},$$ (C.7)

where the potential vector $\mathbf{A}$ of a monochromatic electromagnetic field satisfies  and the gauge $\nabla\cdot\mathbf{A} = 0$; $\mathbf{P} = \sum_i^N \mathbf{p}_i$ is the momentum of electrons that fill the VB at $T$=0K. Then, by applying the time-dependent perturbation theory, in the limit of large time and by using the Dirac delta function definition $\delta(x) = \lim_{a\to\infty}\frac{1}{\pi a}\frac{1-\cos ax}{x^2}$, one obtains the expression of probability rate for absorption in the first order of approximation

$$R = \frac{\pi\,e^2\,E_0^2}{2\,m_0^2\,\hbar\omega^2}\sum_{i,f}\left|M_{if}\right|^2\delta\left(E_{if} - \hbar\omega\right),$$ (C.8)

where $M_{if} = \left\langle\Psi_i\left|\mathbf{e}\cdot\mathbf{P}\right|\Psi_f\right\rangle$ is the optical matrix element and $E_{if} = \hbar\left(\omega_i - \omega_f\right)$. By replacing the Dirac function by the Lorentzian $\delta(x) = \lim_{\gamma\to 0}\frac{1}{\pi}\frac{\gamma}{\gamma^2 + x^2}$ in Eq. (C.8), Eq. (C.6) of the single QD absorption coefficient becomes

$$\alpha_{QD} = \frac{e^2}{cn\varepsilon_0 m_0^2\,\omega V_{QD}}\sum_{i,f}\frac{\gamma\left|M_{if}\right|^2}{\gamma^2 + \left(E_{if} - \hbar\omega\right)^2},$$ (C.9)

where $\gamma$ is the homogeneous electronic broadening.

The excitonic spinless QD Hamiltonian formed by the kinetic part and Coulomb electron-hole interaction in the second quantization is written as in Ref. [19]:

$$H_D = \sum_m E_m^e c_m^+ c_m + \sum_m E_m^h h_m^+ h_m + \sum_{m,n,p,q}V_{mnpq}^{eh} c_m^+ h_n^+ h_p c_q,$$ (C.10)

where the first, second term stands for electron, hole kinetic energy, and the third for electron-hole Coulomb interaction, respectively; $c_m^+ (c_m)$ are the creation (annihilation) electron state $m$ in CB and $h_m^+ (h_m)$ for hole state $m$ in VB. The Coulomb matrix element is

$$V_{mnpq}^{eh} = -e^2 \left(4\pi\varepsilon_0 \bar{\varepsilon}_r\right)^{-1} \iint_V d\mathbf{r}_e d\mathbf{r}_h \psi_m^e(\mathbf{r}_e)^* \psi_n^h(\mathbf{r}_h)^* \left|\mathbf{r}_e - \mathbf{r}_h\right|^{-1} \psi_p^h(\mathbf{r}_h) \psi_q^e(\mathbf{r}_e), \qquad (C.11)$$

with $\bar{\varepsilon}_r$ the screened relative dielectric constant, which holds for a homogenized value. By applying the all bra configurations $\langle 0 | h_j c_i$ to the QD Schrödinger $H_D |\Psi\rangle = E|\Psi\rangle$ one obtains the secular equation

$$\sum_{i,j,k,l=1} \left[ \left(E_i^e + E_j^h - E\right)\delta_{ik}\delta_{jl} + V_{ijlk}^{eh} \right] C_{kl} = 0. \qquad (C.12)$$

By using the series expansion of the Coulomb Green function in spherical harmonics

$$\frac{1}{\left|\mathbf{r}_e - \mathbf{r}_h\right|} = 4\pi \sum_{L=0}^{\infty} \sum_{m=-L}^{L} \frac{1}{2L+1} \frac{r_<^L}{r_>^{L+1}} Y_{Lm}^*(\hat{\mathbf{r}}_e) Y_{Lm}(\hat{\mathbf{r}}_h), \qquad (C.13)$$

the angular integrals of the Coulomb matrix elements $V_{ijlk}^{eh}$ factorize for electron and hole and they are computed analytically by using Gaunt's formula.

Exciton formation implies the initial state is the ground state, consequently the optical matrix element reads $M_{0\alpha} = \langle 0 | \mathbf{e} \cdot \mathbf{P} | \Psi^{(\alpha)} \rangle$. Thus, by writing the transition momentum operator in the basis set $\left\{ |0\rangle, |\Psi^{(\alpha)}\rangle \right\}$ as

$$\mathbf{P} = \sum_\alpha |0\rangle \langle 0 | \sum_{i=1}^N \mathbf{p}_i | \Psi^{(\alpha)} \rangle \langle \Psi^{(\alpha)} | + h.c., \qquad (C.14)$$

one obtains (see Ref. [20])

$$\begin{aligned} M_{0\alpha} &= \mathbf{e} \cdot \sum_{i,j,k=1} C_{ij}^{(\alpha)} \langle 0 | \mathbf{p}_k c_i^+ h_j^+ | 0 \rangle = \mathbf{e} \cdot \sum_{i,j=1} C_{ij}^{(\alpha)} \langle \psi_j^h u_v | \mathbf{p} | \psi_i^e u_c \rangle \\ &= \mathbf{e} \cdot \mathbf{p}_{vc} \sum_{i,j=1} C_{ij}^{(\alpha)} \langle \psi_j^h | \psi_i^e \rangle \end{aligned}, \qquad (C.15)$$

where $\mathbf{p}_{vc} = \langle u_v | \mathbf{p} | u_c \rangle$, and the last equality is obtained by making use of the slow spatial variation of the envelope wave functions over regions of the unit cell size and the orthonormality of the Bloch cell wave functions. By introducing the Kane momentum matrix element, $P = -i(\hbar/m_0)\langle s | p_z | z \rangle = -i(\hbar/m_0)\langle u_c | p_z | u_v \rangle = -i(\hbar/m_0) p_z^{cv}$, with $E_P = 2m_0 |P|^2 / \hbar^2$, and considering the polarization unit vector, $\mathbf{e}$, parallel to the quantization axis, $z$, for example, one obtains

$$|M_{0\alpha}|^2 = \frac{E_P m_0}{2} \left| \sum_{i,j=1} C_{ij}^{(\alpha)} \left\langle \psi_j^h \middle| \psi_i^e \right\rangle \right|^2 , \tag{C.16}$$

and expression of the single QD absorption coefficient from Eq. (C. 9) becomes

$$\alpha_{QD}(\omega) = \frac{\alpha_0}{\omega} \sum_{\tau} \left| \sum_{i,j=1} C_{ij}^{(\tau)} \left\langle \psi_i^e \middle| \psi_j^h \right\rangle \right|^2 \frac{\gamma}{(E_\tau - \hbar\omega)^2 + \gamma^2} , \tag{C.17}$$

where $\alpha_0 = e^2 E_P / (2nc\varepsilon_0 m_0 V_{QD})$. In the applicative part of the main text (see Eq. (5)), we estimate Ep=20.4eV for CdSe/11CdS/ZnS and Ep=19.1eV for ZnTe/6ZnSe/ZnS from Ref. [10].

The *colloidal* absorption coefficient is obtained from a probabilistic hitting target model (the target is the QD) by neglecting the light scattering. We divide the solution volume in a cubic grid (cube edge of $c_{QD}^{-1/3}$ with $c_{QD}$ the concentration of QDs in solution), each cube containing a single QD. For dilute solutions $D << c_{QD}^{-1/3}$ (where $D$ is the effective path length of the light passing through a single QD) and we consider a light beam which propagates perpendicular to the grid cubes surface.

We write the light intensity after it passes through a large number of grid cubes aligned parallel to the direction of light propagation. The light absorbed by the first grid cube is probabilistically calculated as $I_0 \left(1 - e^{-\alpha_{QD}D}\right) \left(2Rc_{QD}^{1/3}\right)^2$, where $\left(2Rc_{QD}^{1/3}\right)^2$ is the probability of the light beam to hit the QD inside the grid cube. After light passes through the first grid cube its intensity is

$$I_1 = I_0 \left[ 1 - \left(1 - e^{-\alpha_{QD}D}\right)\left(2Rc_{QD}^{1/3}\right)^2 \right]. \tag{C.18}$$

Next, we introduce the colloidal absorption coefficient $\alpha$ in solution by

$$I = I_0 e^{-\alpha_{sol} L} , \tag{C.19}$$

where $L$ is the light path length, iterate $N$ times Eq. (C.39), make $L = N c_{QD}^{-1/3}$, compare the iteration result with the light intensity absorption written with Eq. (C.19) and find the absorption coefficient for a dilute solution of colloidal QDs is given by

$$\alpha_{sol} = -c_{QD}^{1/3} \ln\left[ 1 - 4R^2 c_{QD}^{2/3} \left(1 - e^{-\alpha_{QD}D}\right) \right]. \tag{C.20}$$